\documentclass{ws-procs9x6}

\begin{document}

\title{Recent Results on Interstellar Turbulence}

\author{Miguel A. de Avillez$^{1,2}$ and Dieter Breitschwerdt$^{2}$}

\address{$^{1}$Department of Mathematics, University of \'Evora, Portugal\\ 
$^{2}$Institut f\"ur Astronomie, Universit\"at Wien, Austria\\
E-mail: mavillez,breitschwerdt@astro.univie.ac.at}

\maketitle

\abstracts{
The statistical properties of interstellar turbulence are studied by
means of three-dimensional high-resolution HD and MHD simulations of a
SN-driven ISM. It is found that the longitudinal and transverse
turbulent length scales have time averaged (over a period of 50 Myr)
ratios of 0.5-0.6, almost similar to the one expected for isotropic
homogeneous turbulence. The mean characteristic size of the larger
eddies is found to be $\sim 75$ pc. Furthermore, the scalings of the
structure functions measured in the simulated disk show unambiguous
departure from the Kolmogorv (1941) model being consistent with
the latest intermittency studies of supersonic turbulence (Politano \&
Pouquet 1995; Boldyrev 2002). Our results are independent of the
resolution, indicating that convergence has been reached, and that the
unresolved smaller dissipative scales do not feed back on the larger
ones.}

\section{Introduction}

Interstellar turbulence is mainly driven by the energy injected into
the ISM by supernovae, with the driving scale still being uncertain. 
It is also unclear what the statistical properties of the
turbulent interstellar gas are, if the full available range of
energies is taken into account. In this paper we discuss the
statistical properties of the interstellar turbulence in the Galactic
disk based on three-dimensional adaptive mesh refinement (AMR)
simulations of the ISM, which include the disk-halo-disk circulation
[\cite{avi00}]. In particular we explore the injection scales (section
2) and the scalings of the velocity structure functions (section 3) of
the interstellar turbulent gas and discuss (section 4) their implications.

\section{The injection scale of interstellar turbulence}

\begin{figure}[thbp]
\centering
\includegraphics[width=0.3\hsize,angle=-90]{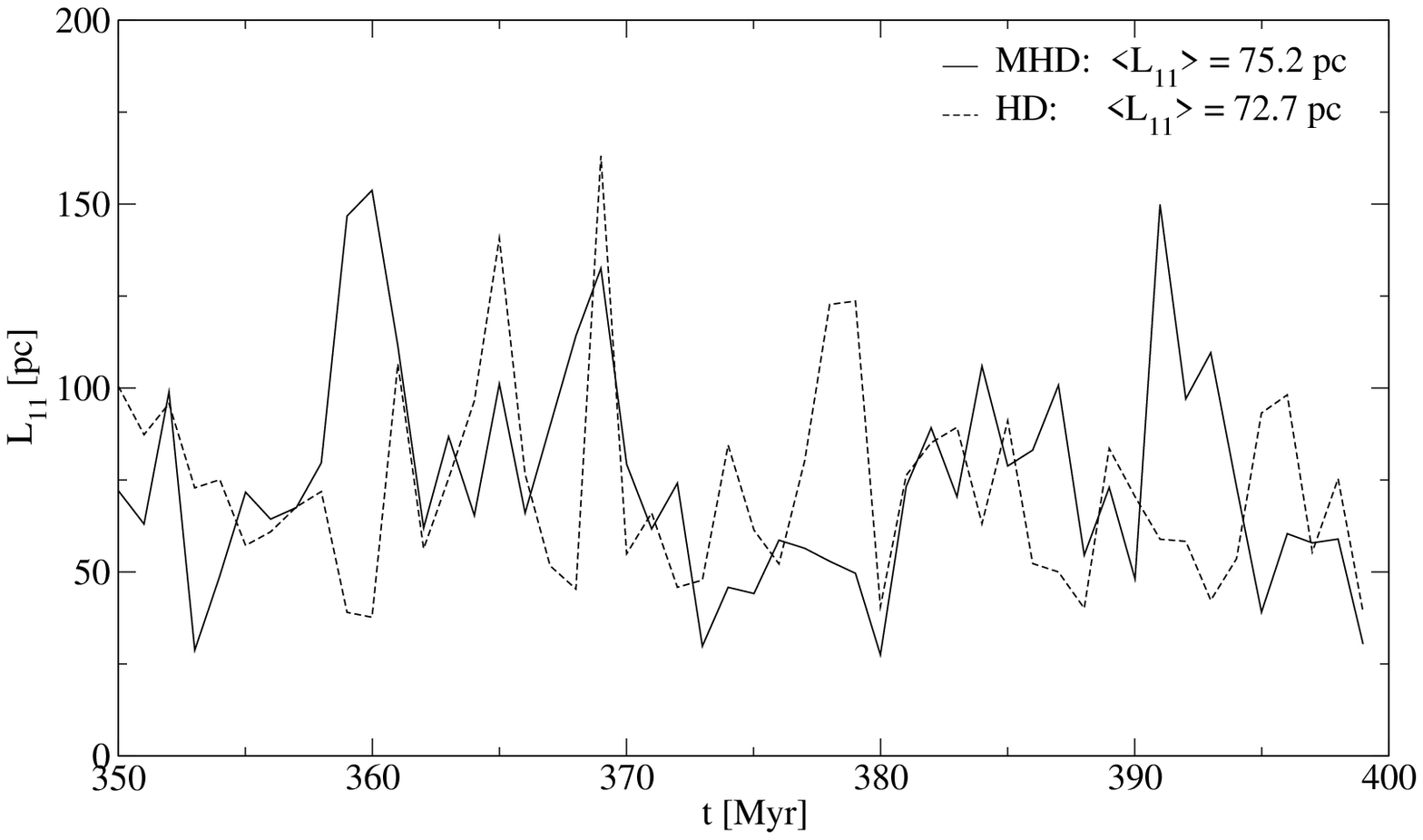}\includegraphics[width=0.3\hsize,angle=-90]{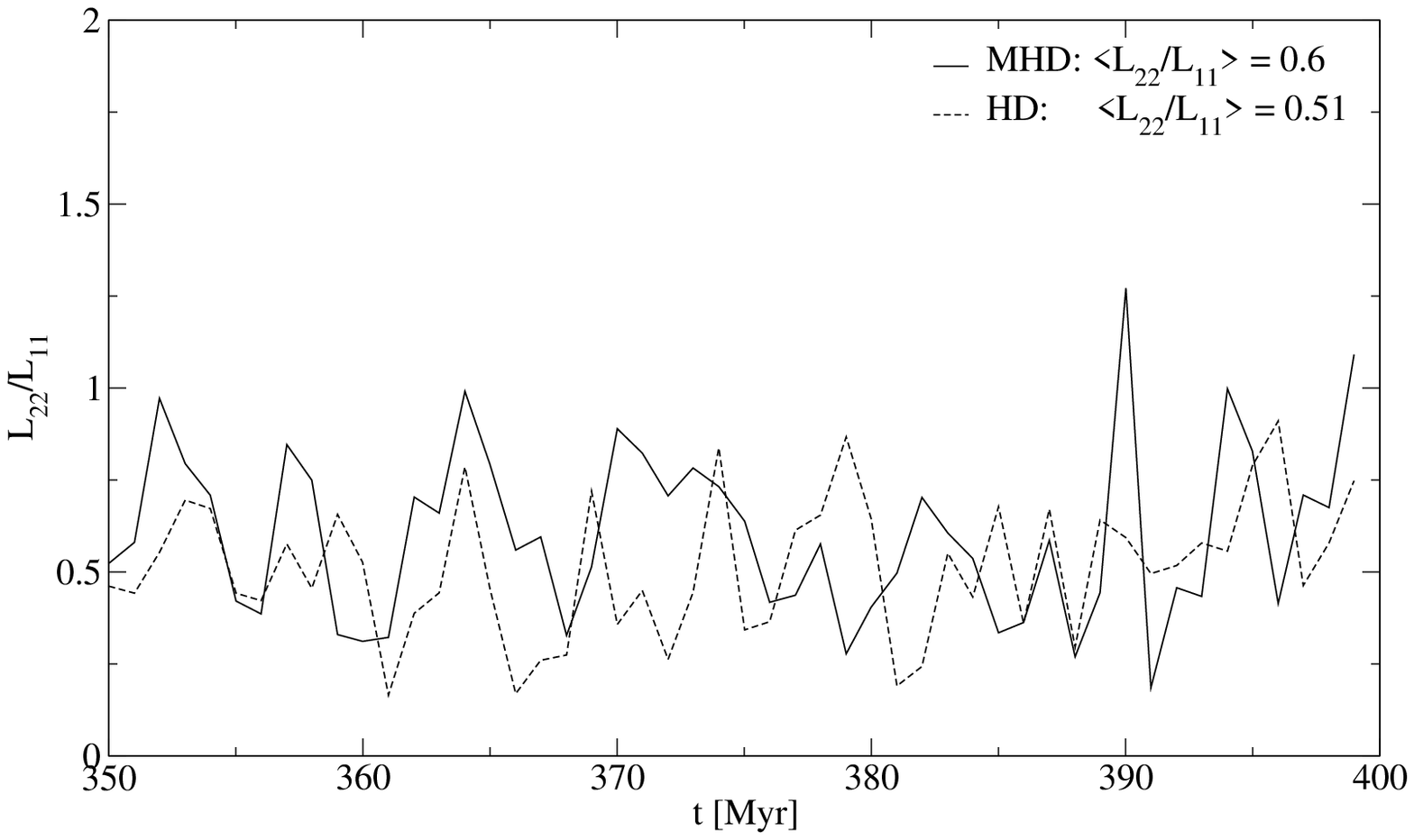}
\caption{History of the characteristic size (given by $L_{11}$) of
the larger eddies (left panel) and of the ratio $L_{22}/L_{11}$
(right panel) for the HD (dashed line) and MHD (solid line) runs.}
\label{l11}
\end{figure}

The outer scale of the turbulent flow in the ISM is related to the
scale at which the energy in blast waves is transferred to the
interstellar gas. Such a scale can be determined by means of the
longitudinal and transversal correlation lengths, $L_{11}$ and
$L_{kk}$, respectively. Here, $k=2,3$ refer to the directions
perpendicular to the 1-direction along which the correlation lengths
are calculated. For isotropic turbulence $L_{kk} = 0.5
L_{11}$. Figure~\ref{l11} shows the history of $L_{11}$ (left panel)
and $L_{22}/L_{11}$ (right panel) during 50 Myr of evolution of the
 unmagnetized and magnetized ISM. For details on how
$L_{11}$ and $L_{22}$ are calculated see [\cite{ab2006}]. Although
$\langle L_{11}\rangle_{t} \sim 75$ pc, the scatter of
$L_{11}$ around its mean results from the oscillations in the local star formation rate, 
where the formation and merging of superbubbles, is responsible for the peaks 
observed in the two plots.
The similarity between the average HD and MHD injection scales is due to the 
fact that magnetic pressure and tension
forces cannot prevent break-up as long as $L\lesssim \beta_{P} \lambda$,
where $L$ and $\lambda$ are the scale lengths of thermal and magnetic
pressures (including tension forces) and $\beta_{P}=4\pi P/B^{2}$ is
the plasma beta.

Despite the large scatter seen in $L_{22}/L_{11}$, the time average
$\langle L_{22}/L_{11}\rangle_{t}$ over the 50 Myr period is 0.51 and
0.6 for the HD and MHD runs, respectively.  The discrepancy from 0.5
by about 20\% in the MHD case is a consequence of the anisotropy introduced by the field into the flow. The
$\langle L_{22}/L_{11} \rangle_{t}\sim 0.5$ in the HD run indicates that in a
statistical sense the interstellar unmagnetized turbulence is roughly
isotropic.

\section{Scalings of the Structure Functions}
\begin{figure}[thbp]
\centering
\includegraphics[width=0.5\hsize,angle=0]{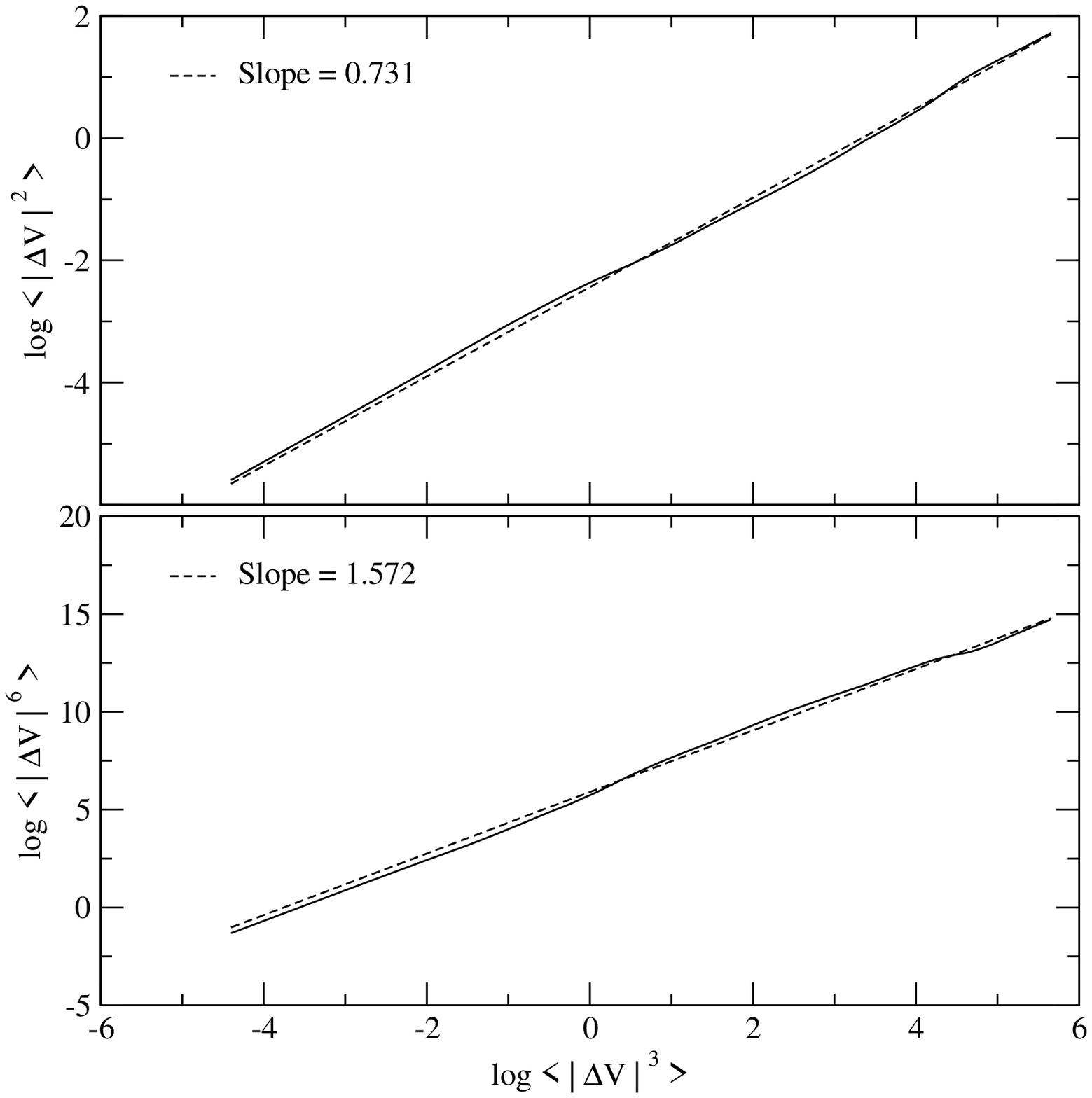}\includegraphics[width=0.5\hsize,angle=0]{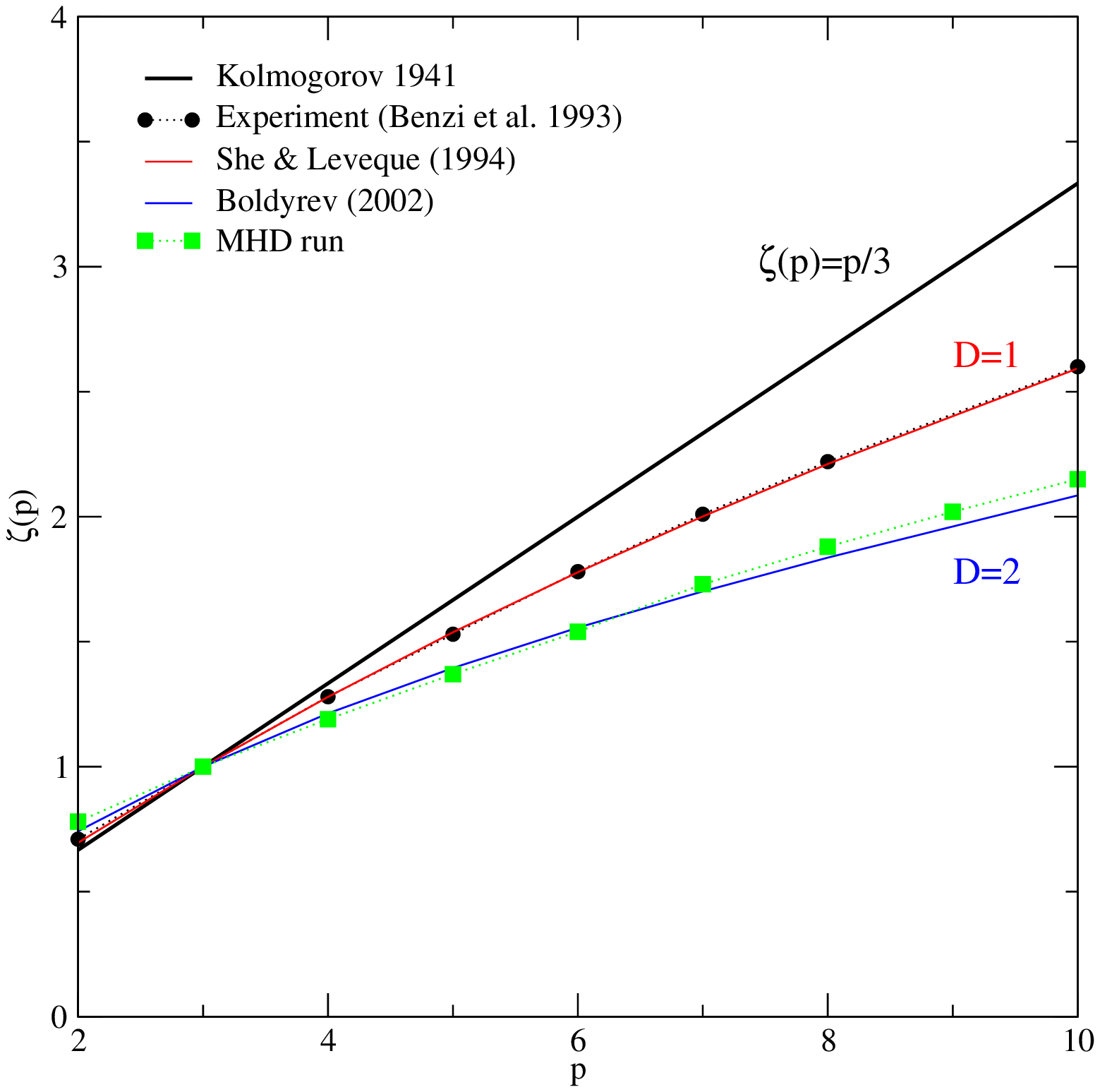}
\caption{\emph{Left:} Log-log plot of $\langle \left|\Delta
V\right|^{2}\rangle$ (top panel) and $\langle \left|\Delta
V\right|^{6}\rangle$ (bottom panel) as function of
$\langle\left|\Delta V\right|^{3}\rangle$ and its best fits in dashed
lines. \emph{Right:} Comparison between theory (plus MHD simulations) and experiments of the exponents $\zeta(p)$ vs. order p.}
\label{slopes}
\end{figure}

The statistics of turbulent flows in physical space is commonly
characterized by the velocity structure functions of order $p$ defined
as $S_{p}(l)=\langle\left|\Delta V_{l}\right|^{p}\rangle$, with
$\Delta V_{l}=v(x+l)-v(x)$, where $v(x+l)$ and $v(x)$ are the
velocities along the $x-$axis at two points separated by a distance
$l$, such $\eta \ll l\ll L$, $L$ being the energy integral scale and
$\eta$ the Kolmogorov scale, respectively. For homogeneous and
isotropic turbulence, the Kolmogorov (K41) [\cite{K41}] theory
predicts that $\langle\left|\Delta V\right|^{3}\rangle=-\frac{4}{5}
\epsilon l$ for $l\gg \nu^{3/4}\epsilon^{-1/4}$. Here, $\nu$ is the
kinematic viscosity and $\langle\;\rangle$ stands for average over the
probability density function of $\Delta V(l)$. As a consequence of
this relation, within the inertial range one can write $\langle
\left|\Delta V\right|^{p}\rangle\propto \langle \left|\Delta
V\right|^{3}\rangle^{\zeta(p)}$. Experiments show that this
expression, with the same scaling exponents, is valid in a wide range
of length scales for large as well as small Reynolds numbers, even if
no inertial range is established [\cite{benzi93}]. Figure~\ref{slopes}
displays, for the unmagnetized interstellar gas, $\langle \left|\Delta
V\right|^{2}\rangle$ (top of left panel) and $\langle \left|\Delta
V\right|^{6}\rangle$ (bottom of left panel) as functions of
$\langle\left|\Delta V\right|^{3}\rangle$ and its best fits in dashed
lines with slopes 0.73 and 1.57 corresponding to the scalings of
$\zeta(2)$ and $\zeta(6)$, respectively. These scalings are different
than those predicted by the K41 theory given by $p/3$ and similar to
those proposed by Boldyrev [\cite{boldyrev2002}].
More generally, the structure functions can be written as
$S_{p}(l)\propto l^{\zeta(p)}$, where $\zeta(p)$ is given by
$\zeta(p)/\zeta(3)=\gamma p +C\left(1-\Sigma^{p}\right)$, with
$C=(1-3\gamma)/\left(1-\Sigma^{3}\right)$ and $\zeta(3)=1$; $\Sigma$
is a measure of the degree of intermittency of the flow, 
$C$ (the Hausdorff codimension of the support of the most singular
dissipative structures) is related to the structures' Hausdorff
dimension $D$ by $C=3-D$. In incompressible flows this 
corresponds to one-dimensional vortex filaments
[\cite{SL94}], hence $C=2$, while the most singular dissipative
structures in supersonic turbulence are two-dimensional shock fronts
[\cite{boldyrev2002}], and in fully developed MHD turbulence these are 2D
micro-current sheets [\cite{politano95}]; hence 
$C=1$. $\gamma$ is related to very high order moments
and has the value of 1/9 for incompressible flows [\cite{SL94}], with a
similar value quoted by [\cite{boldyrev2002,politano95}] for
supersonic HD and MHD turbulence. Hence, $\Sigma^{3}=2/3$ or $1/3$ if
$C=1$ or 2. The degree of intermittency in the compressible turbulence
is higher than in the incompressible case. In Kolmogorov turbulence
$\zeta(p)=p/3$, which implies that $C=0$ independently of the value of
$\Sigma$. A comparison between the different theoretical scalings and
the ones derived from the MHD run is shown in Figure~\ref{slopes}. Although the relative scalings, which express the dependence of the
moments at different orders, are universal, they show unambiguous
departure from the Kolmogorov law due to intermittency of interstellar
turbulence. For further discussion on these results see
[\cite{ab2006}].

\section{Final Remarks}

In subsonic turbulence energy is injected at the outer scales and
transferred to the smallest scales, with dissipation by molecular
viscosity setting in at the Kolmogorov inner scale.  Dissipation is a
passive process as it proceeds at a rate determined by the inviscid
inertial behaviour of the large eddies. Such an energy cascading
corresponds to a divergence-free behaviour of the flow in the inertial
range. However, the ISM is a compressible medium swept up by shocks,
which are in general more efficient in dissipating energy, although
most of it occurs again at small scales through low Mach number
shocks.

This research is supported by Portuguese Science Foundation (FCT)
through projects BSAB-455 to M.A. and PESO/P/PRO/40149/2000 to the
authors. 

\end{document}